\documentclass[runningheads]{llncs}
\usepackage[T1]{fontenc}
\usepackage{hyperref}
\usepackage{comment}
\usepackage{caption}
\usepackage{subcaption}
\usepackage{graphicx}
\usepackage{color}
\usepackage{amsmath}
\usepackage{algorithm}
\usepackage{algpseudocode}

\newcommand\blfootnote[1]{%
 \begin{NoHyper}%
  \renewcommand\thefootnote{}\footnote{#1}%
  \addtocounter{footnote}{-1}%
  \end{NoHyper}%
}

\begin{document}
\title{Annotated Biomedical Video Generation using Denoising Diffusion Probabilistic Models and Flow Fields}
\author{Anonymous}
\authorrunning{Anonymous}
\institute{Anonymous Organization \\
\email{***@***.***}
}

\author{Rüveyda Yilmaz\inst{1} \and
Dennis Eschweiler\inst{1,2,*} \and
Johannes Stegmaier\inst{1}}
\authorrunning{R. Yilmaz et al.}
%
\institute{Institute of Imaging and Computer Vision, RWTH Aachen University, Germany
\email{$\lbrace$rueveyda.yilmaz, johannes.stegmaier$\rbrace$@lfb.rwth-aachen.de}\\
\and
Department of Diagnostic and Interventional Radiology, University Hospital RWTH Aachen, Germany\\
\email{deschweiler@ukaachen.de}
}

\titlerunning\space{BVDM}
\maketitle

\begin{abstract}
The segmentation and tracking of living cells play a vital role within the biomedical domain, particularly in cancer research, drug development, and developmental biology.
These are usually tedious and time-consuming tasks that are traditionally done by biomedical experts. 
Recently, to automatize these processes, deep learning based segmentation and tracking methods have been proposed.
These methods require large-scale datasets and their full potential is constrained by the scarcity of annotated data in the biomedical imaging domain.
To address this limitation, we propose Biomedical Video Diffusion Model (BVDM), capable of generating realistic-looking synthetic microscopy videos.
Trained only on a single real video, BVDM can generate videos of arbitrary length with pixel-level annotations that can be used for training data-hungry models.
It is composed of a denoising diffusion probabilistic model (DDPM) generating high-fidelity synthetic cell microscopy images and a flow prediction model (FPM) predicting the non-rigid transformation between consecutive video frames.
During inference, initially, the DDPM imposes realistic cell textures on synthetic cell masks which are generated based on real data statistics.
The flow prediction model predicts the flow field between consecutive masks and applies that to the DDPM output from the previous time frame to create the next one while keeping temporal consistency.
BVDM outperforms state-of-the-art synthetic live cell microscopy video generation models.
Furthermore, we demonstrate that a sufficiently large synthetic dataset enhances the performance of cell segmentation and tracking models compared to using a limited amount of available real data.

\blfootnote{$^*$Funded by the German Research Foundation DFG (STE2802/2-1).}

\end{abstract}
\section{Introduction}
Segmentation and tracking of living cells have significant importance in disease diagnosis and treatment, and basic research in developmental biology.
Traditional methods employed to handle these tasks are labor-intensive, time-consuming, and inherently subjective, relying heavily on the expertise of biomedical professionals \cite{emami2021computerized}.
With recent advances in deep learning, biomedical research has achieved significant progress in automating these tasks \cite{ronneberger2015u,stringer2021cellpose,scherr2020cell,he2017cell}.
Automatic segmentation algorithms play a fundamental role in the precise identification and delineation of cell boundaries within microscopic images \cite{hollandi2022nucleus}.
These algorithms contribute not only to accelerating the pace of analysis but also to ensuring a higher degree of accuracy and reproducibility in cell characterization.
In cancer research, for instance, where understanding cellular behavior is central to unraveling the complexities of tumor development and progression, accurate segmentation is crucial for identifying abnormal cell morphology, growth patterns, and other critical features indicative of malignancy \cite{emami2021computerized}.
Tracking algorithms, on the other hand, extend the utility of segmentation by enabling the monitoring of individual cells over time \cite{hollandi2022nucleus}.
This temporal dimension is particularly vital in developmental biology, where observing and quantifying dynamic processes such as cell division, migration, and differentiation are essential for deciphering the intricate mechanisms underlying organism development.
In drug development, the ability to track cells through various stages of treatment facilitates the assessment of drug efficacy and potential side effects, aiding researchers in refining therapeutic strategies \cite{kwak2010single}. \\
\textbf{Contributions:}
Despite being fundamental for aforementioned deep learning methods, annotated living cell microscopy data is limited due to privacy regulations, ethical concerns, and the intricacies of data collection in highly specialized and controlled medical environments.
Additionally, the complex and time-consuming nature of manual data annotation contributes to the limited availability of large and diverse datasets.
To eliminate this problem and unlock the full potential of deep learning based segmentation and tracking methods in the biomedical domain, we propose BVDM, capable of generating annotated live cell microscopy videos.
Building upon DDPM-based image synthesis, BVDM generates realistic-looking and temporally consistent videos of arbitrary length with pixel-level annotations.
Specifically, our contributions are three-fold:
(1) Given real cell microscopy videos, BVDM learns the spatiotemporal relations embedded in the data from multiple aspects.
(2) It generates a diverse set of realistic microscopy videos with pixel-level annotations that can be used to train deep learning models;
(3) We demonstrate that the segmentation and tracking performance of these models increases when trained on synthetically generated diverse videos compared to limited real data. \\
\textbf{Related Work:}
To address the challenges posed by data scarcity, various generative algorithms \cite{mercan2020virtual,jose2021generative,eschweiler2024denoising} aimed at synthesizing annotated microscopy image datasets have been developed in recent years.
While significant advances have been made in generating static images, the generation of videos introduces a higher level of complexity due to the temporal dimension.
Here, maintaining textural and structural consistency between consecutive time frames becomes a crucial requirement.
To this end, classical methods \cite{svoboda2016mitogen,mavska2014benchmark,mavska2023cell} have attempted to model characteristics of cell microscopy videos by combining manually designed functions that represent cell texture and shape, the point spread function, and the noise properties of simulated microscopes.
However, the labor-intensive nature of designing these functions for each cell type and different microscopes poses a significant challenge in producing realistic-looking videos.
Recently, GAN-based approaches \cite{bahr2021cellcyclegan,celard2023study} have demonstrated the capability to generate realistic and practically useful fully-annotated microscopy image sequences.
Notably, these approaches alleviate the main challenges associated with classical methods.
More recently, DDPM-based models have demonstrated improved performance in data generation, highlighted by recent studies \cite{muller2022diffusion,muller2023multimodal,dhariwal2021diffusion,oh2023diffmix,wu2023retinal,eschweiler2024denoising} that outperforms previous GAN-based models. \\
\section{Method}
Inspired by the success of DDPMs \cite{ddpm_vanilla} in image generation, we propose BVDM, the first diffusion-based model capable of generating annotated microscopy video datasets.
\begin{figure}[tb]
\includegraphics[width=0.95\textwidth]{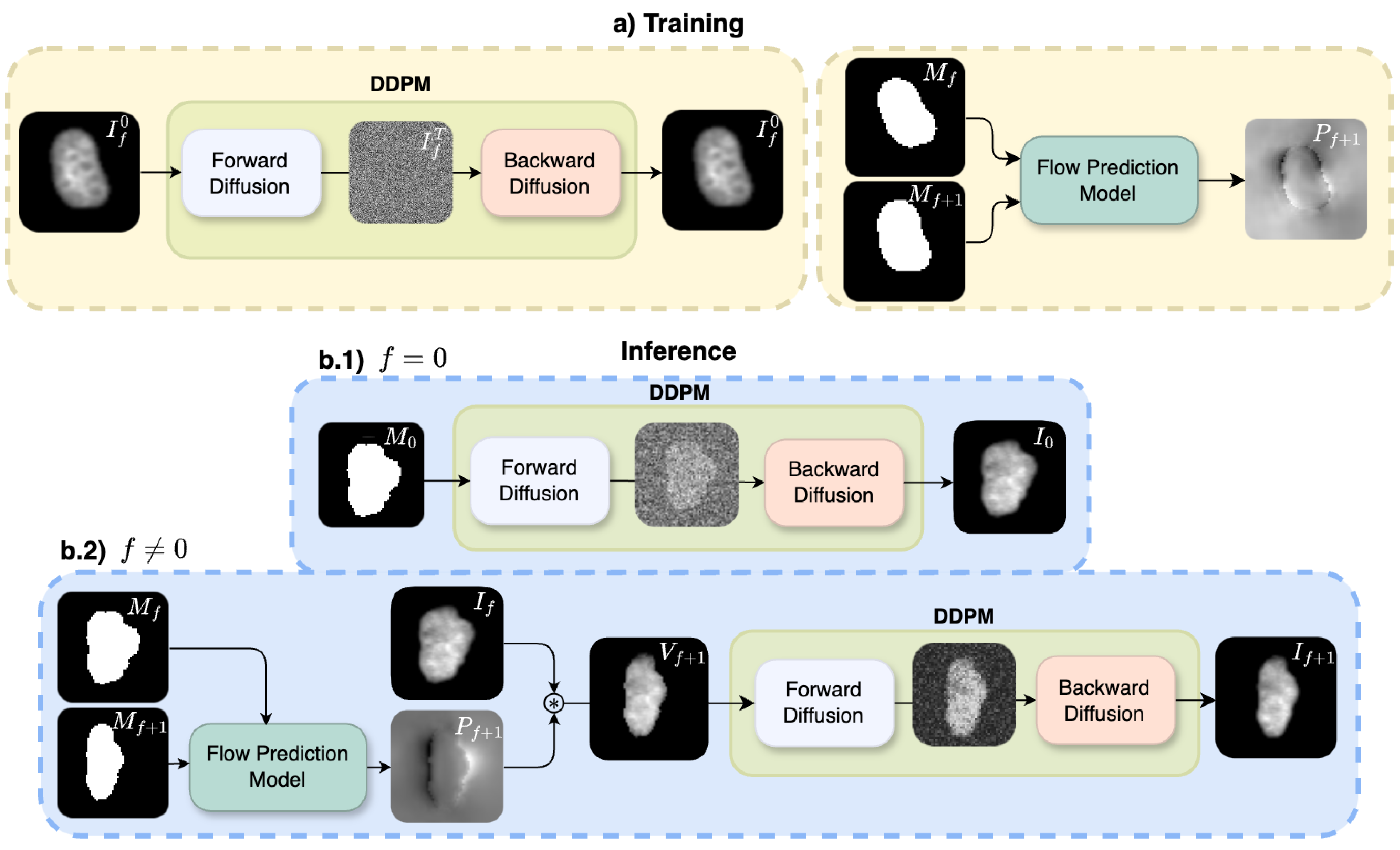}
\caption{Overview of BVDM in training (a) and inference (b.1, b.2). (a) The DDPM and VoxelMorph are trained independently on random images $I_f$ and consecutive masks $M_f$ and $M_{f+1}$ respectively. (b.1) During inference, the DDPM generates the texture for the first appearance of each cell. (b.2) For the other frames, flow field prediction from VoxelMorph is applied to the output from the previous iteration, and the result is fed to the DDPM.}
\label{fig:method}
\end{figure}
Given 2D live cell microscopy videos, BVDM learns the spatiotemporal relations embedded in the data at multiple levels and uses this information to generate realistic live cell microscopy videos that are temporally consistent.
At a basic level, the cell properties are modeled by a statistical shape model \cite{bahr2021cellcyclegan} trained on real data \cite{zhong2012unsupervised}.
This model captures the geometries and average brightness of cells to ensure that realistic cell shapes are generated.
To impose realism on the generated cell shapes, a DDPM is utilized to learn the texture characteristics of the cellular images.
In tandem, an FPM is used to learn relations between time points and to temporally propagate generated cell textures, such that generated videos consistently change over time. \\
\textbf{DDPMs} are a class of generative models capable of generating synthetic images using a diffusion process \cite{sohl2015deep,ddpm_vanilla,eschweiler2024denoising}.
Based on a noise scheduler, a predefined amount of Gaussian noise is added to an image $I^t$ at the diffusion step $t$ until the final image $I^T$ is distributed by $\mathcal{N}(0,\,1)\,$.
This procedure is called the \textit{forward process}, and it is reversed by training a U-Net \cite{ronneberger2015u} to execute the inverse during the \textit{backward process} (Fig.~\ref{fig:method}a).
This trained model is then used at inference time to generate synthetic images by starting from a noisy image and iteratively denoising it (Fig.~\ref{fig:method}b.1). \\
\textbf{FPMs} are used for image registration, aligning different images of the same object acquired at different time steps or from different perspectives.
The goal is to find the flow field, a non-rigid spatial transformation to align two images accurately.
Given a source $I_{src}$ and target image $I_{tgt}$, the model predicts the flow field $g_\theta (I_{src}, I_{tgt})$ to be applied on $I_{src}$ to register it to $I_{tgt}$.
By comparing the intensities of corresponding pixels in the source and target images, the model learns to minimize the dissimilarity between them while ensuring smooth deformations.
In this work, we use VoxelMorph \cite{vxm} as the image registration model for ensuring textural consistency in subsequent video frames. \\ 
\textbf{Training:}
During training, a real annotated live cell microscopy dataset \cite{fluo_dataset} is used and both the DDPM and VoxelMorph are independently trained (Fig. \ref{fig:method}a).
A single image $I_f$ is randomly selected from the dataset and fed to the DDPM for learning to generate realistic images as explained previously.
To train VoxelMorph, a pair of consecutive mask annotation images $M_{f}$, $M_{f+1}$ from the frames $f$ and $f+1$ of a cell are randomly selected and fed into the model as the source $I_{src}$, and the target $I_{tgt}$, respectively.
Then, the model is trained to predict the non-rigid transformation that aligns the masks, which is in this case the temporal flow between two consecutive video frames. \\
\textbf{Inference:}
During inference, we start by randomly generating synthetic masks $M_f$ with $f \in \{0,...,F\}$ for all frames based on the statistical shape model constructed from real data \cite{bahr2021cellcyclegan,zhong2012unsupervised}.
Morphological consistency of the generated consecutive masks is ensured by a Gaussian-based interpolation between different statistical shape models \cite{bahr2021cellcyclegan}.
Next, we generate the first appearance of each cell using only the DDPM, and then generate the following frames iteratively using both the DDPM and the VoxelMorph models.
Unlike \cite{ddpm_vanilla}, which starts from pure noise to generate an image during inference, we perform the forward process also at inference time on our synthetic masks.
However, we stop after $T_{f=0} < T$ diffusion steps to prevent losing all the shape information contained in the mask.
This results in a noisy, yet not completely distorted mask (Fig. \ref{fig:method}b.1).
It guides the DDPM regarding the shape of the cell to be generated in the backward process that is performed using the model trained on real data.
To generate the subsequent frames, we compute the flow field $P_{f+1}$ between the current and the next synthetic masks $M_{f}$ and $M_{f+1}$ using VoxelMorph (Fig. \ref{fig:method}b.2).
Then, using this flow field, we transform the DDPM output $I_{f}$ into the shape of the mask $M_{f+1}$ at frame $f+1$ resulting in $V_{f+1}$.
This step establishes the connection between consecutive synthetic video frames, yielding consistency in texture over time rather than having a collection of independently generated DDPM outputs.  
Inherent in its nature, VoxelMorph tends to smooth out images while performing registration since it applies interpolation to compute values at the image grid.
To prevent this, we feed the VoxelMorph output $V_{f+1}$ to the DDPM for $T_{f\neq0} \ll T_{f=0} <T$ diffusion steps.
We perform this whole process iteratively for $F$ frames to generate a video ($I_0$,...,$I_F$) of a single cell (see Fig. ~\ref{fig:method}).
Finally, multiple cell videos are placed randomly in a bigger scene with random translation, rotation, and movement to form synthetic live cell microscopy data.
\section{Experiments}
\textbf{Dataset:}
In our experiments, we use real microscopy sequences featuring HeLa cells \cite{fluo_dataset} obtained from the cell tracking challenge \cite{mavska2023cell}.
The dataset comprises two grayscale videos, each with a resolution of $700\times1100$ and a length of 92 frames.
In our experiments, we use one video for training and the other one for validation.
Within each video, multiple cells are present per frame, and our initial step involves extracting each cell into a separate image.
This extraction is essential as BVDM is trained on individual cell images.
We rely on the silver truth annotations provided with the dataset to locate the cells during this extraction process.
Subsequently, we create multiple centered cell images and the corresponding mask images with dimensions of $96\times96$. \\
\textbf{Evaluation Method:}
We assess the performance of BVDM through a comprehensive set of metrics, including segmentation accuracy (SEG) \cite{mavska2014benchmark}, tracking accuracy (TRA) \cite{tra}, Fréchet Video Distance (FVD) \cite{fvd}, and Fréchet Inception Distance (FID) \cite{fid}.
The FID score measures the disparity between the features of real and synthetic images, while the FVD score performs a comparable analysis on videos, both evaluating the realism of the synthetic data \cite{fid,fvd}.
SEG is a widely used metric for evaluating segmentation methods, akin to the intersection over union score.
Similarly, TRA, designed to evaluate tracking algorithms, quantifies the minimum modifications needed to align predicted cell tracks with ground truth tracks in a given sequence.
Given our focus on enhancing the performance of segmentation and tracking algorithms through synthetic cell videos, SEG and TRA are the primary evaluation metrics.
Specifically, we train cell segmentation and tracking models on synthetic videos and assess their performance on real videos using SEG and TRA. \\
\textbf{Implementation Details:}
To perform these evaluations, we employ the tracking algorithm from \cite{ben2022graph}, currently the top-performing model in the cell tracking challenge.
As a segmentation model we employ \cite{segmentation} since \cite{ben2022graph} also uses that for detecting cells during inference to further perform tracking.
For the DDPM, we configured the batch size to 32, set the learning rate to $5 \times 10^{-4}$, and conducted training for 10000 iterations.
\begin{table}[tb]
\begin{center}
\captionof{table}{Comparative evaluations with SEG, TRA, FVD, and FID for real, MitoGen \cite{svoboda2016mitogen}, CellCycleGAN \cite{bahr2021cellcyclegan} and BVDM data. 
\label{table:comparisons}}
\begin{tabular}{lccccc}
\hline 
Data set & SEG $\uparrow$ & TRA $\uparrow$ & FVD $\downarrow$ & FID $\downarrow$ \\
\hline 
Real & 0.823 &  0.976 & - & - \\
MitoGen & 0.822 & 0.972 & 84.6 & 49.6 \\
CellCycleGAN & 0.781 & 0.953 & 350.4 & 70.7 \\
Ours & \textbf{0.829} & \textbf{0.979} & \textbf{79.4} & \textbf{29.6}  \\
\hline 
\end{tabular}
\end{center}
\end{table}
\begin{figure}[h]
\centering
\includegraphics[width=0.8\textwidth]{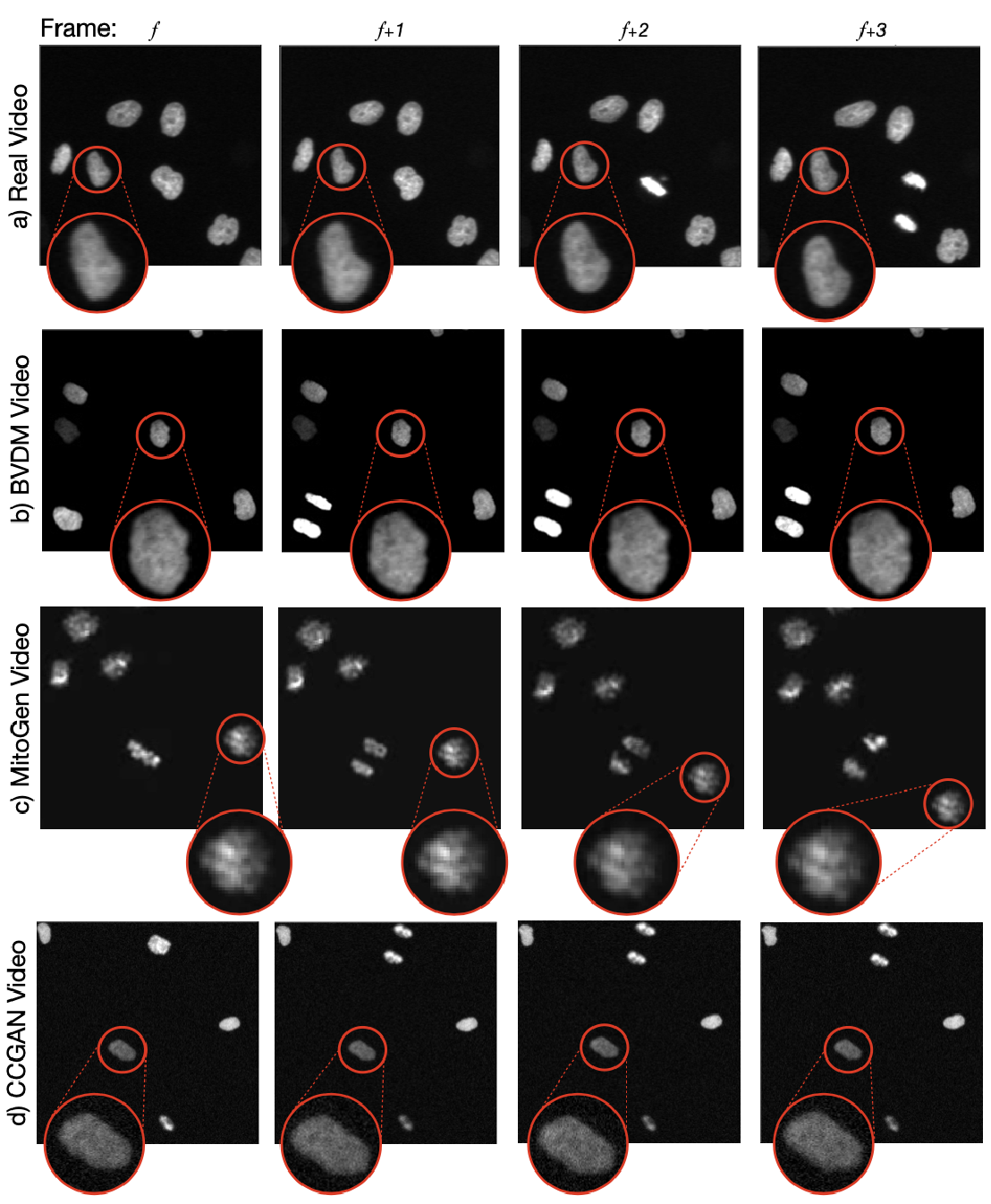}
\caption{Qualitative comparisons of the texture consistency across frames from the real dataset \cite{mavska2014benchmark} (a), BVDM (b), MitoGen \cite{svoboda2016mitogen} (c), and CellCycleGAN \cite{bahr2021cellcyclegan} (d).}
\label{fig:qualitative}
\end{figure}
Additionally, the flow prediction model was trained with a batch size of 32, a learning rate of $10^{-4}$, for 2000 iterations using silver truth annotations from the dataset.
To train our models, we used an NVIDIA GeForce RTX 3090 GPU. \\
\textbf{Experimental Results:} 
In our experiments, we train the synthetic video generation models BVDM, MitoGen \cite{svoboda2016mitogen} and CellCycleGAN \cite{bahr2021cellcyclegan} on the real training sequence and generate two synthetic video sequences for each method.
Then, we train the segmentation and tracking models on these synthetic video sequences and evaluate the performance on the real validation sequence (Table \ref{table:comparisons}). 
Notably, BVDM outperforms the other methods in all four metrics.
Furthermore, the segmentation and the tracking models trained on synthetic BVDM data achieve an SEG of 0.829 and a TRA of 0.979 surpassing even the tracking model trained on the real training sequence. 
This suggests that training models on multiple synthetic BVDM sequences generated based on a single real video leads to superior performance compared to using the real video itself. 
This improvement implies that BVDM augments the data by generating diversified examples that are characteristically similar to the real dataset.
Using such synthetic data proves especially valuable in scenarios where annotated data is scarce, a common challenge in biomedical imaging.
In Figure \ref{fig:qualitative}, we qualitatively present consecutive example video frames from the real and synthetic datasets that confirm the obtained results.
The close-up views show that MitoGen data exhibits realistic cell shapes but lacks sufficient variations in cell textures at different time points. 
On the other hand, CellCycleGAN successfully introduces these variations but lacks textural similarity to the real data.
BVDM, however, satisfies both of the requirements, leading to superior scores. \\
\textbf{Ablation Studies:} 
We conducted ablation experiments to optimize the values of $T_{f=0}$ and $T_{f \neq 0}$ with respect to the evaluation metrics.  
The results for the most successful combinations are presented in Tables \ref{table:comparison_t1} and \ref{table:comparison_t2}.
In each table, one parameter is fixed while the other is varied, with $T_{f \neq 0} = 10$ for Table \ref{table:comparison_t1} and $T_{f=0} = 200$ for Table \ref{table:comparison_t2}.
Generally, both the SEG and TRA metrics tend to decrease as the values of $T_{f=0}$ and $T_{f\neq0}$ increase.
This can be attributed to the fact that an increased number of diffusion time steps introduces added Gaussian noise during the forward process, thereby degrading the shape defined by the input mask and leading to a discrepancy between the synthesized image and the corresponding mask.
This discrepancy misleads the segmentation and tracking models by providing faulty ground truth annotations during training. 
On the other hand, when $T_{f=0} = 100$ in Table \ref{table:comparison_t1} both SEG and TRA scores are lower compared to $T_{f=0} = 200$.
This is because fewer diffusion steps result in synthesized cell textures that appear more artificial, adversely affecting the performance of segmentation and tracking models when applied to real data.
Regarding FVD and FID, we observe better scores for increasing values of $T_{f=0}$.
This is attributed to the improved realism of cell textures as more diffusion steps are taken.
Unlike SEG and TRA, FVD and FID are indifferent to the degree of match between the shapes of generated cells and their synthetic annotation masks. 
As a result, their values do not increase for higher values of $T_{f=0}$.
For $T_{f=0}=200$, we explored various values of $T_{f\neq0}$, detailed in Table \ref{table:comparison_t2}. 
In the last row of the table, the flow field is not predicted between consecutive frames; instead, each frame is independently generated with 200 diffusion time steps.
This was done to observe what happens when images are generated independently, without ensuring texture consistency across consecutive frames.
With increasing values of $T_{f\neq0}$, segmentation and tracking models exhibit decreasing performance, and FVD is higher in general. 
The reduction in performance can be attributed to the increasing noise introduced during the forward process, leading to a loss of structures on the synthetic cell texture from the previous frame and a subsequent decrease in temporal consistency, reflected in poorer TRA and FVD scores.
The decline in SEG follows a similar trend due to the mismatch between annotations and generated images for higher diffusion time steps. 
Conversely, FID is minimized for $T_{f \neq 0} = 200$, as it operates solely at the image level and does not account for the temporal consistency of consecutive frames.
When $T_{f \neq 0}$ is close to 0, all scores are worse contrary to the common trend.
This is because the flow field prediction model applies interpolation to pixel values when transforming the images, resulting in unrealistically smooth textures.
Considering the scores from Tables \ref{table:comparison_t1} and \ref{table:comparison_t2}, the combination $T_{f=0} = 200$ and $T_{f\neq 0} = 10$ emerges as the optimal parameter combination. 
\begin{table}[tb]
\caption{$T_{f=0}$ and $T_{f\neq 0}$ are the number of diffusion time steps taken while generating the first and the later frames of a cell respectively.}
\centering
\begin{subtable}{.45\textwidth}
\centering
\begin{tabular}{l|ccccc}
\hline 
$T_{f=0}$ & SEG $\uparrow$ & TRA $\uparrow$ & FVD $\downarrow$ & FID $\downarrow$ \\
\hline 
100 & 0.822 & 0.967 & 82.9 & 32.3 \\
200 & \textbf{0.829} & \textbf{0.979} & 79.4 & 29.6 \\
400 & 0.815 & 0.960 & 75.4 & 28.0 \\
600 & 0.786 & 0.947 & \textbf{74.7} & \textbf{20.7} \\
\hline
\end{tabular}
\caption{Quantitative results for varying values of $T_{f=0}$ for $T_{f\neq 0}$=10. \label{table:comparison_t1}}
\end{subtable}
\begin{subtable}{.45\textwidth}
\centering
\begin{tabular}{l|ccccc}
\hline 
$T_{f\neq 0}$ & SEG $\uparrow$ & TRA $\uparrow$ & FVD $\downarrow$ & FID $\downarrow$ \\
\hline 
0 & 0.820 & 0.972 & 83.9 & 35.6 \\
10 & \textbf{0.829} & \textbf{0.979} & \textbf{79.4} & 29.6 \\
30 & 0.827 & 0.970 & 81.3 & 29.1 \\
50 & 0.821 & 0.965 & 83.7 & 29.4 \\
200 & 0.813 & 0.941 & 91.4 & \textbf{28.4} \\
\hline
\end{tabular}
\caption{Quantitative results for varying values of $T_{f\neq 0}$ for $T_{f=0}$=200. \label{table:comparison_t2}}
\end{subtable}
\end{table}
Although not all metrics align on this choice, it is essential to note that our primary motivation for proposing BVDM was to enhance the performance of segmentation and tracking algorithms.
Therefore, it is desirable to select parameter values that maximize SEG and TRA scores, given that FVD and FID scores for this combination are not significantly different from the best achieved scores.
Additionally, opting for lower values of $T_{f = 0}$ and $T_{f \neq 0}$ alleviates the computational load as the number of diffusion steps in the backward process is the primary bottleneck. 
The data generation process is approximately 42.7\% faster when using $T_{f=0} = 200$ and $T_{f\neq 0} = 10$ compared to the configuration with $T_{f=0} = 200$ and $T_{f\neq 0} = 200$.
\section{Conclusion}
Leveraging DDPMs and flow prediction, BVDM proposes a novel approach for generating synthetic live cell microscopy videos, outperforming the current state-of-the-art in biomedical video generation.
It tackles the scarcity of annotated real biomedical datasets by generating temporally consistent videos with pixel-level annotations.
Using our synthetic data for training even demonstrates superior performance compared to training exclusively on a limited amount of real data.
Our code will be available at \url{https://github.com/ruveydayilmaz0/BVDM}.

\bibliographystyle{splncs04}
\bibliography{egbib}
\newpage
\textbf{Supplementary Material}

\begin{figure}[h]
\centering
\includegraphics[width=0.8\textwidth]{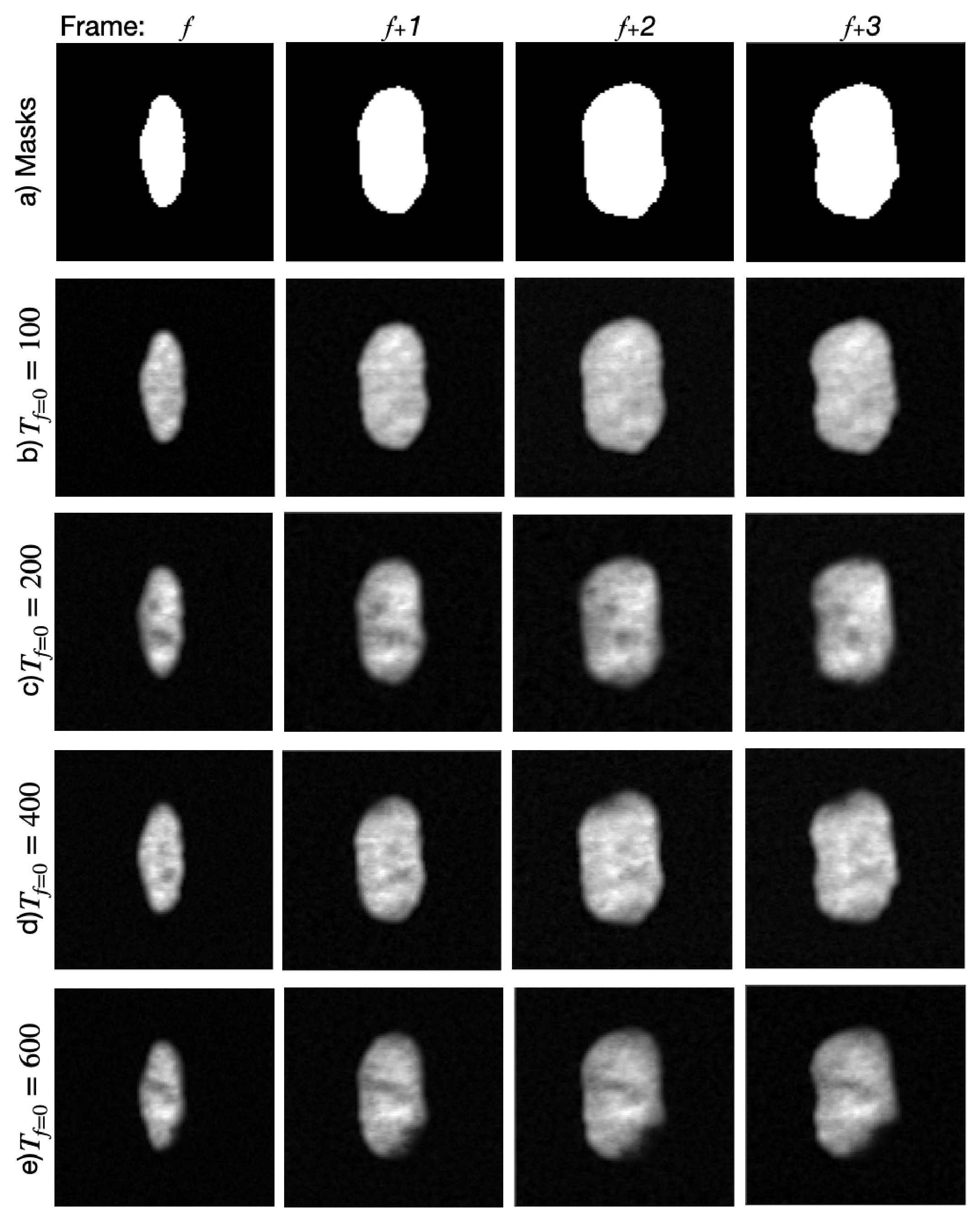}
\caption{Sample masks from four consecutive frames (a) and qualitative results for $T_{f=0}=100$ (b), $T_{f=0}=200$ (c), $T_{f=0}=400$ (d), and $T_{f=0}=600$ (e) when $T_{f\neq 0}=10$ as a demonstration of the mismatch between the mask annotations and the generated images depending on the diffusion time step $T_{f=0}$ taken.}
\label{fig:fig1}
\end{figure}

\begin{figure}[h]
\centering
\includegraphics[width=0.8\textwidth]{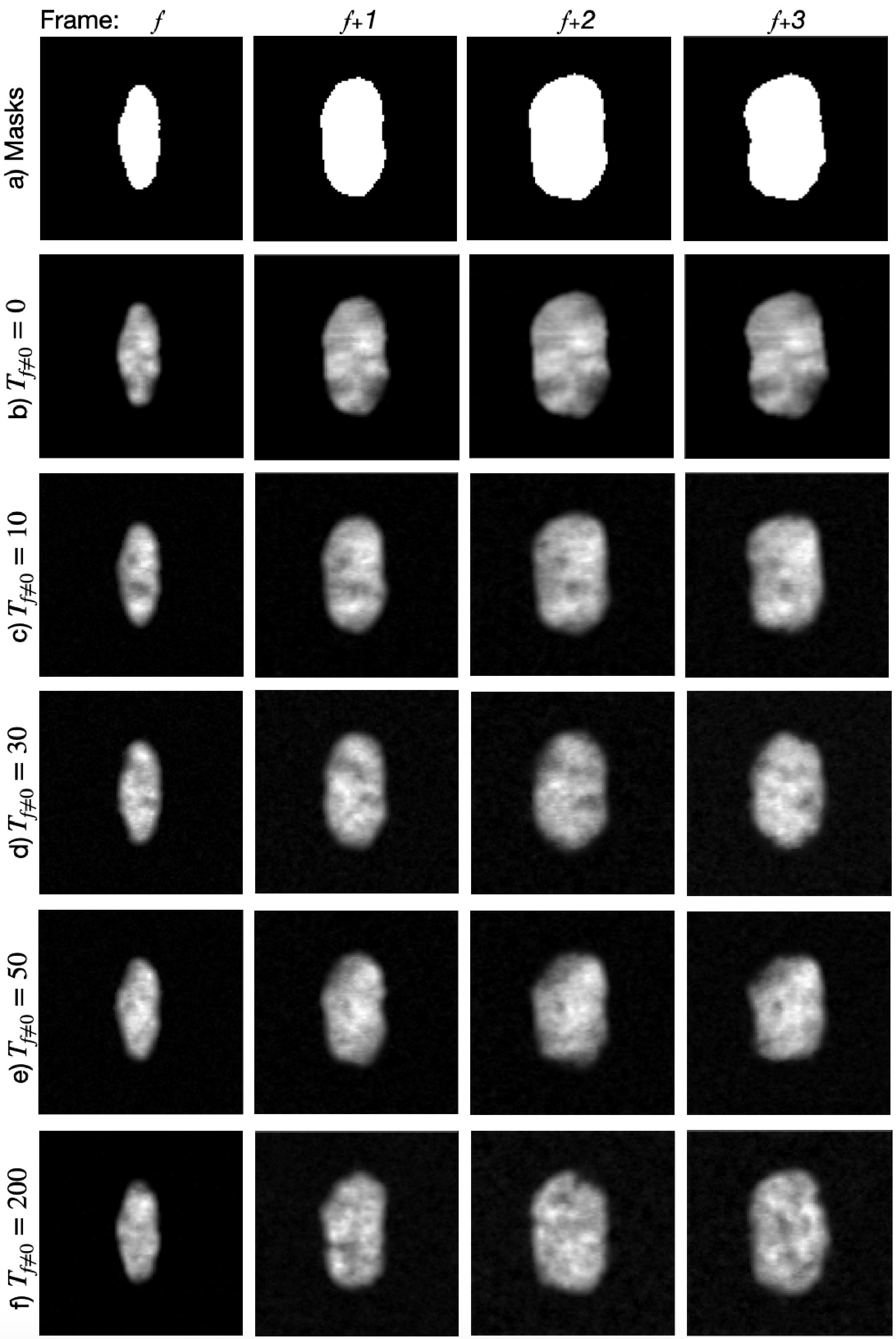}
\caption{Sample masks from four consecutive frames (a) and qualitative results for $T_{f\neq 0}=0$ (b), $T_{f\neq 0}=10$ (c), $T_{f\neq 0}=30$ (d), $T_{f\neq 0}=50$ (e), and $T_{f\neq 0}=200$ (f) when $T_{f=0}=200$ as a demonstration of the mismatch between the mask annotations and the generated images, and the texture consistency depending on the diffusion time step $T_{f\neq 0}$. For (f), the flow field is not predicted between consecutive frames; instead, each frame is independently generated with 200 diffusion time steps.}
\label{fig:fig2}
\end{figure}

\end{document}